# Comparative study of the relationships between CO isotopic luminosities and infrared luminosity for the Galactic dense cores

Yan Sun    Yu Gao

(Purple Mountain Observatory, Chinese Academy of Science, 2 West Beijing Road, Nanjing, Jiangsu 210008, China)

**Abstract** Combining the $^{12}CO(1\text{-}0)$, $^{13}CO(1\text{-}0)$, and $C^{18}O(1\text{-}0)$ data with IRAS four band data, we here estimate the physical parameters such as size, viral mass, and CO J=1-0 isotopic and infrared luminosities for 29 dense molecular clouds from two published CO samples. We further analyze the various correlations between CO J=1-0 isotopic luminosities and infrared luminosity (star formation rate, SFR) and discuss the relationships between the molecular gas tracers and SFR. The results show that $^{12}CO(1\text{-}0)$, $^{13}CO(1\text{-}0)$ and $C^{18}O(1\text{-}0)$ luminosities have tight correlations with each other. CO J=1-0 isotopic luminosities and SFR show weak correlations with lager scatter than the HCN-IR correlations of 47 dense cores in the Galaxy and 65 external star-forming galaxies. This might be interpreted as that both the SFR and star formation efficiency are mainly determined by the molecular gas at high volume density rather than high column density.

The collapse of giant molecular clouds (GMCs) will initiate star formation, and the dense gas in the dense cores of molecular clouds will constantly be converted into stars. Star formation rate (SFR) and star formation efficiency (SFE) of the molecular clouds are essential for the understanding of the star formation environment and process[1]. Thus, the SFR and SFE have been the concern of astronomers for a long time. SFR is expressed in terms of the total mass of stars that formed in the molecular cloud per unit time, and can be traced by multi-wavelength emission of enormous objects, like the bright star, supernova, recombination lines (H$_\alpha$, H$_\beta$), UV emission, radio continuum emission in HII region and far infrared (IR) emission, etc[2]. SFE is expressed in terms of the SFR per unit mass of molecular gas, SFR/M(H$_2$). By comparing different methods in measuring

SFR, Kennicutt (1998) concluded that infrared luminosity is the best SFR indicator when the dust heating is dominated by young stellar objects and high extinction is encountered in the active star forming regions [3].

The star formation (SF) law describes the relation between SFR and the density of the star-forming gas content. The Schmidt law was firstly studied in the Galaxy and expressed as SFR$\propto \rho^{\alpha}$, where $\rho$ is the density of atomic gas, and $\alpha$ is a constant in the range of 1-3[4]. However, we know that star formation occurs in molecular clouds and does not have any direct relation with atomic gas. Since 1970, the molecular line CO emission was detected in the ISM, particularly the CO detection in external galaxies; molecular gas ($H_2$), which is an important component of the ISM has been included in the raw material for star formation by astronomers. Thus, the SF law has been adopted globally to galaxies in a new expression $\sum_{SFR} = A \sum_{gas}^{N}$, where $\sum_{SFR}$ is the disk averaged surface SFR, $\Sigma_{gas}$ is the surface density of atomic gas and molecular gas, N is the power-law index, and A is the absolute SFR [5].

Recently, rapid progress in the study of the SF law in galaxies has been made. For 61 normal spiral galaxies and 36 starburst galaxies, Kennicutt (1998) showed that the disk averaged SFRs that were derived from H$_\alpha$ and infrared luminosities are correlated with the total surface gas density derived from HI and CO, with a power-law index of $1.4\pm0.15$[5]. But, this index is strongly dependent on the property of the sample as the index is quite different for the normal spiral galaxies and starburst galaxies. $^{12}$CO is more likely a total molecular gas tracer and is not closely related with star formation, since stars are born in the much denser part of molecular clouds. To better understand the true physical conditions of star formation, it is essential to observe cold and dense molecular gas. HCN is one of the most important dense gas tracers with a high dipole-moment. For 9 ultraluminous infrared galaxies, 22 luminous infrared galaxies, and 34 normal spiral galaxies, Gao & Solomon (2004) found a tight linear correlation between IR and HCN luminosities with a power-law index of 1.0 [6,7]. This result shows that the SFRs indicated by infrared luminosity strongly depend on the amount of dense gas traced by HCN emission rather than the total molecular gas traced by CO, let alone the neutral atomic gas or the sum of atomic gas and molecular gas (total gas). Recent studies show that most of the molecular gas will not be converted into stars, but rather only a small fraction of dense molecular gas will be converted into stars.

More recently, the multi-band high resolution data have been used in a number of comprehensive studies to constrain the form of the Schmidt law in nearby galaxies on sub-kpc scales (e.g., Kennicutt et al. 2007; Bigiel et al. 2008)[8, 9]. On the scale of 0.5-2kpc in NGC 5194 (M51a), Kennicutt et al. (2007) found that the resolved and extinction corrected SFR versus total gas surface density relation is well represented by a Schmidt power law with index N in the range of 1.37 to 1.56[8], which is similar to the disk averaged result[5]. Nevertheless, at sub-kpc resolution in a sample of 18 nearby galaxies, Bigiel et al. (2008) found a Schmidt power law with index N=1.0$\pm$0.2 relating the resolved surface SFR and total gas surface density[9]. Therefore, discrepancy in the power law index even in the resolved local SF law still exists.

So far, the global SF law on the Galactic GMC scale is rarely studied. It is fundamental to study the global star formation law on such scale. On the GMC scale, the correlations between the various dense molecular gas tracers and IR luminosity deserve to be well studied. Indeed, understanding the relation on this SF scale may shed light on the SF nature of other galaxies, even high-z galaxies. Wu et al. (2005) found a tight linear correlation between IR and HCN luminosities in Galactic GMCs at a much smaller scale[10], which suggests that the global SF law derived on the entire galaxy scale is still valid even on the Galactic GMC scale.

In the sample of Mooney & Solomon (1988), the CO and IR luminosities of 55 GMCs show weaker correlation with larger scatter than the HCN-IR correlation in the dense core sample of Wu et al.[11]. The critical densities of $^{13}CO$ and $C^{18}O$ are higher than that of $^{12}CO$. So we expect that the luminosities of $^{13}CO$ and $C^{18}O$ are also correlated with IR luminosity. Besides, $C^{18}O$, as distinct from CO, CS and HCN, is always optically thin. $^{13}CO$ is also optically thin in most conditions[12]. In other words, both $^{13}CO$ and $C^{18}O$ can trace the regions with much higher column density. Therefore combining the CO J=1-0 isotopic and IRAS four band data, we can estimate the physical parameters such as the size, viral mass, CO isotopic and infrared luminosities for 29 dense cores in GMCs. This paper analyzes the correlations between CO isotopic luminosities and infrared luminosity, and further compares it with other dense gas tracers (like HCN).

1. The sample and the physical parameters of GMCs
1.1 The sample and data

The sample is drawn from two published papers on observations of massive dense Galactic cores that have been fully mapped in CO J=1-0 isotopes[12, 13]. The sample of Sun & Gao (2009) contained 29 GMCs [13], among them DR21S and W75(OH) were too close (about 3') to be distinguished as two separate cores based on the limited resolution of the CO observation [12]. Moreover, the two cores are also unresolved based on the IRAS data, so we regard the two cores as one in this paper to estimate their physical parameters. S235N contains two resolvable cores which are marked as S235NE and S235NW and are discussed respectively later. Finally, this sample still includes 29 GMCs with heliocentric distances less than 4 kpc. Among them, the CO data of 15 dense cores are from Zhang & Gao (2008). The CO isotopic observations were carried out using the 13.7 m radio telescope of Purple Mountain Observatory (PMO) in Delingha, China. The CO isotopes were observed simultaneously and were mapped with beam spacing in steps of 1'. The position switch mode was used. The typical mapping size was 6'×6'. The detailed descriptions of our sample and observations were given in the papers of Zhang & Gao (2008) and Sun & Gao (2009)[12, 13]. The data were processed with Gildas and IDL software.

1.2 The IR luminosity of GMCs.

We use the fluxes in the IRAS point source catalog to estimate the IR luminosity for 27 GMCs which contain the IRAS point sources. If there is more than one IRAS point source in the GMC, we choose the brightest one to estimate the IR luminosity. Aperture photometry is also adopted in the improved reprocessing of the IRAS survey (IRIS) image[14] of two GMCs (W75(OH)/DR21S and W75N) which do not contain the IRAS point source that can be used to derive the IR luminosity. The aperture size is equivalent to the size of the GMC mapped in CO isotopes. Finally, the total IR luminosity (8-1000μm) in terms of the IRAS four band fluxes can be expressed as[15],

$$L_{IR} = 0.56 \times D^2 \times (13.48 \times f_{12} + 5.16 \times f_{25} + 2.58 \times f_{60} + f_{100}),$$

where $f_x$ is the flux of the x band in units of Jy, D is the distance in units of kpc; the total IR luminosity $L_{IR}$ is in units of $L_\odot$. The derived IR luminosity, with L(min)= $1.3 \times 10^3 L_\odot$, L(max)= $1.8 \times 10^6$ $L_\odot$, L(mean)= $1.4\times10^5$ $L_\odot$, and L(median)= $1.3\times10^4$ $L_\odot$, is consistent with the typical IR luminosity of GMCs. Compared with the sample of Mooney & Solomon (1988) and Wu et al. (2005), our sample is dominated by low luminosity GMCs. Table 1-1 lists the distance, the flux of the associated IRAS point source or the extended IRAS source and the derived IR luminosity of 29 GMCs, respectively.

Table1-1 The IR luminosity of GMCs

| Core Name | Distance(kpc) | Fluxes(Jy) | | | | $L_{IR}(L_\odot)$ |
| | | $12\mu$m | $25\mu$m | $60\mu$m | $100\mu$m | $10^4 L_\odot$ |
| (1) | (2) | (3) | (4) | (5) | (6) | (7) |
| CEP-A | 0.73 | 1.124e+01 | 8.203e+02 | 1.283e+04 | 2.047e+04 | 1.73 |
| S231 | 2.3 | 5.609e+00 | 7.466e+01 | 7.223e+02 | 1.310e+03 | 1.08 |
| S88 | 2.1 | 9.316e+01 | 1.185e+03 | 8.686e+03 | 1.321e+04 | 10.62 |
| NGC7538 | 2.8 | 2.427e+02 | 1.781e+03 | 7.073e+03 | 1.414e+04 | 19.69 |
| S106 | 0.6 | 2.045e+02 | 2.510e+03 | 1.014e+04 | 1.313e+04 | 1.11 |
| IRAS19410 | 2.2 | 1.443e+01 | 1.088e+02 | 9.825e+02 | 1.631e+03 | 1.33 |
| IRAS20126 | 1.7 | 2.532e+00 | 1.089e+02 | 1.382e+03 | 1.947e+03 | 0.99 |
| G192 | 2.0 | 1.260e+00 | 6.309e+01 | 4.207e+02 | 5.275e+02 | 0.44 |
| AFGL4029 | 2.2 | 1.991e+01 | 2.118e+02 | 7.679e+02 | 1.083e+03 | 1.20 |
| AFGL5142 | 1.8 | 1.260e+00 | 6.309e+01 | 4.207e+02 | 5.275e+02 | 0.35 |
| S255 | 1.3 | 1.072e+02 | 3.716e+02 | 3.145e+03 | 5.285e+03 | 1.59 |
| G123.07-6.31 | 2.2 | 1.791e+00 | 2.113e+01 | 3.566e+02 | 6.850e+02 | 0.47 |
| S87 | 1.9 | 4.742e+01 | 4.255e+02 | 3.447e+03 | 5.174e+03 | 3.42 |
| S235 | 1.6 | 2.825e+01 | 2.264e+02 | 1.709e+03 | 1.635e+03 | 1.09 |
| S252A | 1.5 | 1.560e+01 | 7.684e+01 | 1.032e+03 | 1.715e+03 | 0.63 |
| W3OH | 1.95 | 4.058e+01 | 5.363e+02 | 9.269e+03 | 1.060e+04 | 8.06 |
| W44 | 3.7 | 1.402e+02 | 1.106e+03 | 1.150e+04 | 3.246e+04 | 53.46 |
| S235NE | 1.6 | 2.607e+01 | 5.687e+01 | 5.934e+02 | 1.465e+03 | 0.52 |
| S235NW | 1.6 | 2.527e+00 | 9.465e+00 | 6.626e+01 | 1.465e+03 | 0.25 |
| G20.08-0.13 | 3.4 | 7.246e+00 | 7.611e+01 | 1.437e+03 | 2.761e+03 | 4.50 |
| OH43.80-0.13 | 2.7 | 5.301e+00 | 1.291e+02 | 1.725e+03 | 2.744e+03 | 3.24 |
| S76E | 2.1 | 3.182e+01 | 2.423e+02 | 2.351e+03 | 4.218e+03 | 2.95 |
| G19.61-023 | 4.0 | 4.784e+01 | 4.069e+02 | 4.635e+03 | 7.093e+03 | 19.53 |
| G35.20-0.74 | 3.3 | 2.730e+00 | 3.038e+00 | 8.810e+00 | 1.408e+02 | 0.13 |
| G35.58-0.03 | 3.5 | 6.012e+00 | 7.707e+01 | 1.507e+03 | 2.594e+03 | 4.78 |
| G24.49-0.04 | 3.5 | 5.143e+00 | 5.940e+00 | 1.874e+02 | 7.271e+02 | 0.90 |
| BFS11-B | 2.0 | 1.478e+01 | 7.896e+01 | 6.877e+02 | 1.215e+03 | 0.81 |
| W75(OH)/Dr21S | 3.0 | 8.174e+02 | 8.442e+03 | 7.131e+04 | 1.381e+05 | 189.84 |
| W75N | 3.0 | 2.471e+02 | 3.178e+03 | 3.581e+04 | 4.066e+04 | 77.77 |

1.3  The physical parameters of GMCs and CO isotopic luminosities

The dense core is defined as the region within the contour of half peak intensity. The size of the core is characterized by the nominal core radius after beam deconvolution. R, which is the radius of a circle that has the same area as the dense core given by,

$$R = D \times (\frac{A_{1/2}}{\pi} - \frac{\theta_{mb}^2}{4})^{1/2},$$

where D is the distance, $\theta_{mb}$ is the main beam size, and $A_{1/2}$ is the measured area within the contour of half peak intensity. The definition of the dense core indicates that the size of the dense core is always larger than the size of the beam $A_{1/2} \geq 3\theta_{mb}^2$. Actually, this definition is more appropriate for the sources with spherically symmetric structure than for the sources with multi-component or filament structures. For the $^{12}CO$ cores, R is in the range of 0.45-4.5 pc, with a mean value of 2.10 pc and median value of 1.86 pc; For the $^{13}CO$ cores, R is in the range of 0.41-4.43 pc，with a mean value of 1.73 pc and median value of 1.37 pc; For the $C^{18}O$ cores, R is in the range of 0.32-3.07 pc, with mean value of 1.24 pc and median value of 1.05 pc. It is obvious that the size of the $^{13}CO$ core is larger than that of the $C^{18}O$ core.

If we assume a spherical system with uniform temperature and density, and disregard the rotation, magnitude and external force, the viral mass ($M_{vir}$) of the dense core can be easily derived [16],

$$\frac{M_{vir}}{M_\odot} = 210 \times (\frac{R}{pc})(\frac{\Delta V}{kms^{-1}})^2,$$

where $\Delta V$ is the width of the CO line at the position of peak intensity. $\Delta V$ reflects the total velocity dispersion which might be caused by turbulence, rotation and inflow, etc., in the process of star formation. The derived viral mass of $^{13}CO$ cores is in the range of $6.89 \times 10^2$-$6.57 \times 10^4$ $M_\odot$, with a mean value of $1.04 \times 10^4$ $M_\odot$ and median value of $3.02 \times 10^3$ $M_\odot$; The derived viral mass of $C^{18}O$ cores is in the range of $3.36 \times 10^2$-$4.39 \times 10^4$ $M_\odot$, with a mean value of $5.28 \times 10^3$ $M_\odot$ and median value of $1.52 \times 10^3$ $M_\odot$.

If we assume a Gaussian brightness distribution for the source and a Gaussian beam, the luminosities of $^{12}CO$, $^{13}CO$ and $C^{18}O$ can be expressed as [10, 11],

$$L_{CO} = 23.5 \times 10^6 \times D^2 \times (\frac{\pi \times \theta_s^2}{4\ln 2}) \times (\frac{\theta_s^2 + \theta_{mb}^2}{\theta_s^2}) \times \int T_R dv,$$

where D is the distance in units of kpc，$\theta_{mb}$ and $\theta_s$ are the angular size of the main beam and the core, respectively, (defined by the CO isotopes) in units of arcsecond, and $\int T_R dv$ is the integrated intensity of CO isotopes in units of Kkms$^{-1}$ which has been corrected by the antenna efficiency. Some GMCs (G20.08-0.13, OH43.80-0.13, G19.61-023, G35.20-0.74, G35.58-0.03, and G24.49-0.04) show multi-component $^{12}CO$ emissions along the line of sight, which are difficult to distinguish from each other. In this case, we use the $^{13}CO$ emission range to estimate the $^{12}CO$ emission range. Finally, the derived CO isotopic luminosities are in units of Kkms$^{-1}$pc$^2$. This method can also be applied to estimate the luminosities of HCN and CS etc. Finally, the derived luminosity of the $^{12}$CO core is in the range of $0.80\times10^2$-$1.60\times10^4$ Kkm s$^{-1}$pc$^2$, with a mean value of $2.31\times10^3$ Kkms$^{-1}$pc$^2$ and median value of $1.27\times10^3$ Kkms$^{-1}$pc$^2$; The luminosity of the $^{13}$CO core is in the range of $0.19\times10^2$-$3.03\times10^3$ Kkms$^{-1}$pc$^2$, with a mean value of $6.03\times10^2$ Kkms$^{-1}$pc$^2$ and median value of $2.40\times10^2$ Kkms$^{-1}$pc$^2$; The luminosity of the C$^{18}$O core is in the range of 2.4-428.0 Kkms$^{-1}$pc$^2$, with a mean value of 57.0 Kkms$^{-1}$pc$^2$ and median value of 13.0 Kkms$^{-1}$pc$^2$. For the sample of Wu et al., the luminosity of the HCN is in the range of 0.4-8000 Kkm s$^{-1}$pc$^2$, with median value of 80 Kkms$^{-1}$pc$^2$ [10], which is more comparable to the luminosity of C$^{18}$O. As a whole, our sample is dominated by low luminosity nearby GMCs.

Table 1-2 lists some parameters of $^{12}$CO, $^{13}$CO and C$^{18}$O emissions, such as $\Delta V$: the width of CO isotopes at the position of peak intensity, $\int T_R dv$: the mean integrated intensity within $A_{1/2}$, R: the radius of the dense core, M$_{vir}$: the viral mass of the $^{13}$CO and C$^{18}$O dense core, and L: the total luminosities of dense cores. Six GMCs (S235, W44, G19.61-023, G35.20-0.74, G35.58-0.03, and G24.49-0.04) have not been fully mapped in $^{12}$CO. $^{12}$CO emission regions are the most extended, whereas C$^{18}$O emission regions are the most compact among $^{12}$CO, $^{13}$CO and C$^{18}$O. Thus, only two cores were not fully mapped in $^{13}$CO and all of the cores were fully mapped in C$^{18}$O. We only give the lower limits of the luminosities for cores which were not fully mapped.

Table 1-2. Physical parameters of dense cores

| Core Name | $\Delta V$ (km s$^{-1}$) | | | $\int T_R dv^a$ (K km s$^{-1}$) | | | $R^b$ (pc) | | | M$_{vir}^c$ (10$^3 M_\odot$) | | L$_{CO}^d$ (10K km s$^{-1}$ pc$^2$) | | |
|---|---|---|---|---|---|---|---|---|---|---|---|---|---|---|
| | $^{12}$CO | $^{13}$CO | C$^{18}$O | $^{12}$CO | $^{13}$CO | C$^{18}$O | $^{12}$CO | $^{13}$CO | C$^{18}$O | $^{13}$CO | C$^{18}$O | $^{12}$CO | $^{13}$CO | C$^{18}$O |
| (1) | (2) | (3) | (4) | (5) | (6) | (7) | (8) | (9) | (10) | (11) | (12) | (13) | (14) | (15) |
| CEP-A | 5.69 | 3.67 | 2.72 | 120.71 | 49.68 | 13.59 | 0.67 | 0.51 | 0.32 | 1.45 | 0.50 | 25.02 | 6.17 | 0.69 |
| S231 | 6.30 | 3.46 | 3.67 | 77.18 | 21.35 | 2.46 | 2.45 | 2.01 | 1.56 | 5.05 | 4.40 | 212.39 | 39.92 | 2.82 |
| S88 | 5.43 | 3.35 | 2.69 | 156.08 | 41.54 | 3.27 | 1.08 | 1.28 | 1.11 | 3.02 | 1.69 | 87.99 | 32.45 | 1.95 |
| NGC7538 | 11.21 | 6.85 | 5.92 | 203.28 | 70.41 | 8.04 | 4.16 | 3.06 | 2.32 | 30.17 | 17.06 | 1603.67 | 303.46 | 20.15 |
| S106 | 5.75 | 2.99 | 2.25 | 136.85 | 40.40 | 5.81 | 0.78 | 0.64 | 0.32 | 1.21 | 0.34 | 37.93 | 7.71 | 0.28 |
| IRAS19410 | 6.81 | 3.09 | 2.35 | 123.46 | 37.99 | 4.03 | 1.51 | 1.18 | 1.06 | 2.37 | 1.23 | 131.53 | 25.59 | 2.19 |
| IRAS20216 | 5.09 | 2.83 | 2.46 | 68.42 | 18.70 | 3.12 | 0.45 | 0.41 | 0.35 | 0.69 | 0.44 | 7.97 | 1.86 | 0.24 |
| G192 | 3.60 | 2.24 | 2.11 | 43.86 | 15.15 | 2.12 | 1.00 | 0.74 | 0.62 | 0.77 | 0.58 | 21.35 | 4.19 | 0.43 |
| AFGL4029 | 4.34 | 2.38 | 2.20 | 84.35 | 22.19 | 2.65 | 1.03 | 0.98 | 0.76 | 1.17 | 0.77 | 43.91 | 10.57 | 0.79 |
| AFGL5142 | 4.95 | 2.92 | 2.75 | 61.91 | 23.45 | 3.14 | 0.86 | 0.66 | 0.64 | 1.17 | 1.01 | 22.40 | 5.18 | 0.65 |
| S255 | 4.34 | 2.91 | 2.87 | 79.83 | 26.09 | 2.70 | 0.95 | 0.73 | 0.70 | 1.29 | 1.21 | 33.79 | 6.59 | 0.64 |
| G123.07-6.31 | 4.25 | 2.09 | 1.80 | 39.84 | 12.74 | 2.02 | 1.86 | 1.57 | 1.03 | 1.45 | 0.70 | 64.12 | 14.81 | 1.04 |
| S87 | 4.22 | 3.54 | 3.40 | 92.66 | 39.48 | 3.29 | 1.23 | 1.02 | 0.64 | 2.69 | 1.55 | 66.31 | 19.85 | 0.70 |
| S235 | 4.59 | 2.79 | 2.39 | 88.13 | 30.01 | 2.90 | 1.41 | 0.92 | 0.65 | 1.50 | 0.77 | 80.66 | 12.00 | 0.61 |
| S252A | 5.91 | 3.28 | 2.44 | 121.67 | 55.83 | 5.45 | 1.70 | 1.38 | 1.12 | 3.12 | 1.39 | 161.37 | 49.42 | 3.17 |
| W3(OH) | 7.07 | 4.69 | 4.16 | 126.53 | 32.20 | 3.40 | 1.47 | 1.26 | 0.83 | 5.79 | 3.03 | 126.99 | 23.98 | 1.17 |
| W44 | 11.73 | 5.80 | 5.20 | 126.69 | 63.08 | 19.48 | 3.86 | 3.08 | 2.63 | 21.74 | 14.96 | 870.94 | 225.78 | 42.21 |
| S235NE | 4.45 | 2.77 | 1.78 | 95.81 | 21.01 | 2.07 | 1.96 | 1.37 | 0.83 | 2.22 | 0.55 | 168.75 | 18.43 | 0.69 |
| S235NW | 4.51 | 2.47 | 2.83 | 93.24 | 20.83 | 2.32 | 1.87 | 1.20 | 0.90 | 1.54 | 1.52 | 149.31 | 14.06 | 0.90 |
| G20.08-0.13 | 10.00 | 4.40 | 3.91 | 31.98 | 10.02 | 1.71 | 3.37 | 1.75 | 1.23 | 7.10 | 3.95 | 167.89 | 14.80 | 1.34 |
| OH43.80-0.13 | 8.09 | 6.08 | 5.30 | 16.93 | 10.94 | 1.93 | 1.67 | 0.75 | 0.62 | 5.80 | 3.66 | 22.37 | 3.40 | 0.45 |
| S76E | 8.20 | 7.00 | 6.31 | 37.62 | 20.95 | 4.41 | 2.50 | 2.30 | 1.75 | 23.67 | 14.64 | 107.72 | 51.00 | 6.31 |
| G19.61-023 | 11.78 | 8.40 | 5.79 | 45.12 | 17.82 | 2.91 | 4.50 | 4.43 | 2.08 | 65.68 | 14.62 | 419.76 | 161.00 | 6.07 |
| G35.20-0.74 | 6.36 | 4.53 | 1.43 | 45.17 | 14.23 | 2.03 | 3.91 | 3.91 | 1.67 | 16.83 | 0.72 | 317.05 | 100.04 | 2.75 |
| G35.58-0.03 | 11.00 | 6.21 | 4.48 | 57.95 | 30.49 | 2.98 | 4.15 | 3.10 | 2.34 | 25.15 | 9.86 | 457.33 | 136.05 | 7.72 |
| G24.49-0.04 | 13.26 | 8.85 | 8.26 | 95.81 | 41.92 | 9.79 | 3.36 | 3.37 | 3.07 | 55.43 | 43.94 | 498.68 | 220.09 | 42.80 |
| BFS11-B | 3.78 | 1.90 | 1.60 | 55.76 | 16.26 | 2.22 | 2.18 | 2.01 | 1.41 | 1.52 | 0.75 | 121.47 | 30.23 | 2.08 |
| W75(OH)/Dr21S | 6.50 | 3.50 | 2.97 | 125.60 | 42.07 | 5.04 | 2.68 | 2.75 | 2.07 | 7.08 | 3.83 | 419.18 | 147.30 | 10.12 |
| W75N | 7.80 | 3.84 | 3.47 | 114.23 | 44.98 | 5.45 | 2.22 | 1.73 | 1.33 | 5.34 | 3.36 | 263.92 | 63.95 | 4.79 |

$^a$ The mean integrated intensity within the half peak contour $A_{1/2}$ after the main beam efficiency correction; S235, W44, G19.61-023 and G35.20-0.74 were not fully mapped in $^{12}$CO, G35.58-0.03 and G24.49-0.04 were not fully mapped in $^{12}$CO and $^{13}$CO, the limit values were given for those sources;
$^b$ The radius of a circle that has the same area as the dense core; limit values of R($^{12}$CO) and R($^{13}$CO) are given for the dense cores mentioned above;
$^c$ Assuming a spherical symmetry system with uniform temperature and density for the dense core, and regardless of the rotation, magnitude and external force
$^d$ The CO isotopic luminosities, the limit values of L($^{12}$CO) and L($^{13}$CO) were given for the dense cores mentioned above;

2. The correlations between $^{12}$CO, $^{13}$CO, and C$^{18}$O luminosities and infrared luminosity

As optically thick ($^{12}$CO) and thin ($^{13}$CO and C$^{18}$O) molecular gas tracers, the luminosities of CO isotopes are well correlated with each other. The two panels of Figure 1 show the almost linear correlations between $^{12}$CO and $^{13}$CO luminosities and between $^{12}$CO and C$^{18}$O luminosities in the log-log plot, respectively. The least-squares fit results are also given in the figure. Some limits in CO isotopic luminosities are indicated with arrows in Figure 1 (2 $^{13}$CO limits and 6 $^{12}$CO limits), and hereafter, in later figures, the limits are simply indicated by cross marks. The correlation between $^{13}$CO and C$^{18}$O luminosities is also tight, which has a larger scatter than the L($^{12}$CO)-L($^{13}$CO) relation but a smaller scatter than the L($^{12}$CO)-L(C$^{18}$O) relation. Therefore, we only give the least-squares fit result for the L($^{13}$CO)-L(C$^{18}$O) relation,

$$\log L_{C^{18}O} = 0.95 \times \log(L_{^{13}CO}) - 1.01, (c.c. = 0.94)$$

To reveal the true physical explanation for the tight correlations among the luminosities of $^{12}$CO, $^{13}$CO and C$^{18}$O, it is essential to remove the other possible factors in the selection effects such as distance and size of GMCs etc. The luminosities of $^{12}$CO, $^{13}$CO and C$^{18}$O, normalized by IR luminosity, also show tight correlations. Figure 2 shows distance independent L($^{12}$CO)/L$_{IR}$-L($^{13}$CO)/L$_{IR}$ and L($^{12}$CO)/L$_{IR}$-L(C$^{18}$O)/L$_{IR}$ relations in the log-log plot. The luminosities of $^{13}$CO and C$^{18}$O, normalized by IR luminosity, also show tight correlation, but the plot is not presented in this paper. The least-squares fit result is given below

$$\log(L_{C^{18}O}/L_{IR}) = 0.93 \times \log(L_{^{13}CO}/L_{IR}) - 1.27, (c.c. = 0.95)$$

The two panels of Figure 3 show the distance and size independent correlations among L($^{12}$CO), L($^{13}$CO) and L(C$^{18}$O) normalized by the areas of dense cores in each CO isotope in the log-log plot, respectively. The correlation between the luminosity surface densities of $^{13}$CO and C$^{18}$O is not presented, and only the least-squares fit result is given below,

$$\log(L_{C^{18}O}/R(C^{18}O)^2) = 0.88 \times \log(L_{^{13}CO}/R(^{13}CO)^2) - 0.61, (c.c. = 0.81)$$

Normalized by the area of dense cores, the correlations among CO isotopes obviously depart from a linear correlation and show large scatters. In Figure 3, the luminosity surface density of $^{12}$CO covers one order of magnitude and shows the largest range, whereas the luminosity surface density of C$^{18}$O has the smallest range among all the CO isotopes. Actually, the luminosity surface density is almost proportional to the brightness temperature, $T_R = T_{bg}e^{-\tau} + (1-e^{-\tau})T_{ex}$. If the brightness temperature of the background is weak ($T_{ex} \gg T_{bg}$), and in the optically thick case, the brightness temperature $T_R \sim T_{ex}$; In the optically thin case for which $\tau < 1$, we have simply $T_R \sim \tau T_{ex}$. In fact, the excitation temperatures of $^{12}$CO, $^{13}$CO and C$^{18}$O are similar within the same dense cores. Because $^{12}$CO is an optically thick molecular line, the range of $^{12}$CO luminosity surface densities just reflects that the excitation temperature varied in different molecular clouds. $^{13}$CO is optically thinner than $^{12}$CO, meanwhile C$^{18}$O is much optically thinner than $^{13}$CO. Therefore, in addition to excitation temperature, the optical depth will also affect the range of luminosity surface densities of $^{13}$CO and C$^{18}$O. These optical depth effects can explain why the luminosity surface density of $^{12}$CO covers the most extended range, whereas the luminosity surface density of C$^{18}$O covers the smallest range. Among the correlations of $^{12}$CO, $^{13}$CO and C$^{18}$O luminosity surface densities, the $^{12}$CO-C$^{18}$O correlation is the weakest. All of those might suggest that the tight correlations among CO isotopic luminosities in Figure 1 are significantly determined by the size of GMC, the excitation temperature and the optical depth. Further more, another obvious result in Figures 1-3 is that $^{12}$CO-$^{13}$CO shows the tightest correlation, whereas $^{12}$CO-C$^{18}$O shows the weakest correlation. This might suggest that in most GMCs, $^{12}$CO and $^{13}$CO indicate a more similar molecular gas structure due to the optically thick nature, whereas C$^{18}$O only penetrates deeply into cloud cores and traces the inner dense region of the GMCs.

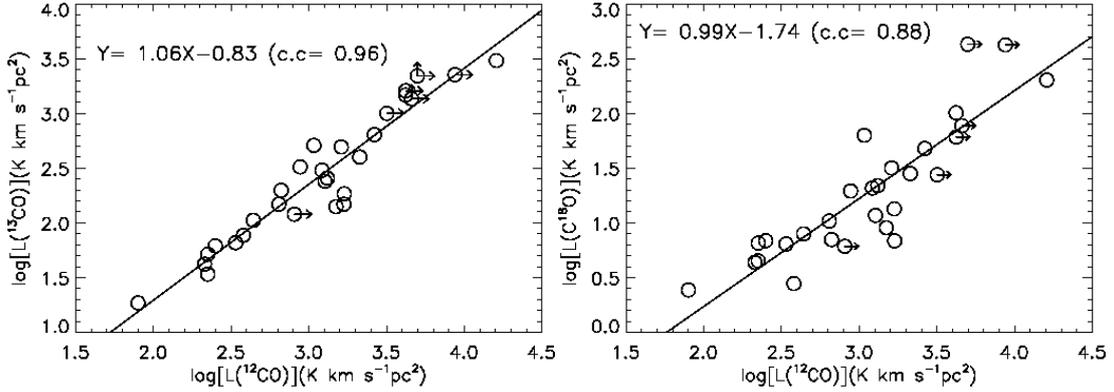

Figure 1. Correlations of $^{12}$CO-$^{13}$CO (left) luminosities and $^{12}$CO-C$^{18}$O (right) luminosities in the log-log plots. The solid lines are the least-squares fit results. The fit results and correlation coefficients are noted in each plot.

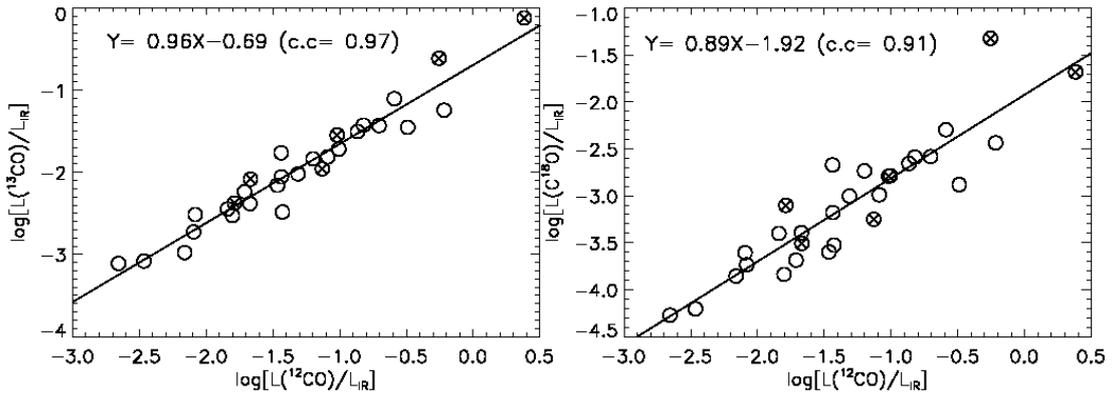

Figure 2. Correlations of $^{12}$CO-$^{13}$CO (left) luminosities and $^{12}$CO-C$^{18}$O (right) luminosities normalized by IR luminosity in the log-log plots. The solid lines are the least-squares fit results. The fit results and correlation coefficients are noted in each plot.

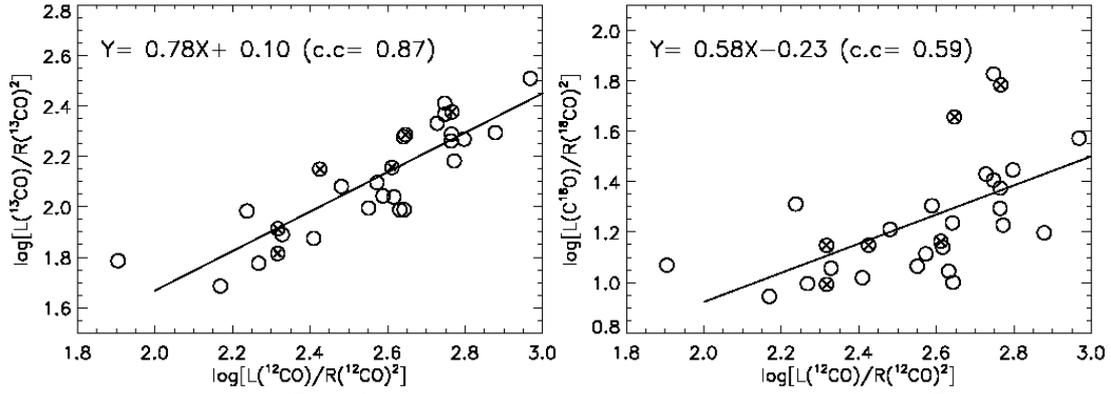

Figure 3. Correlations of $^{12}$CO-$^{13}$CO (left) luminosity surface densities and $^{12}$CO-C$^{18}$O (right) luminosity surface densities in the log-log plots. The size scales of dense cores, R($^{12}$CO), R($^{13}$CO) and R(C$^{18}$O), are referred to Table 1-2.

The mass of molecular gas can be indicated by CO isotopic luminosities meanwhile the SFR can be indicated by the IR luminosity which is derived from IRAS four band data. Figure 4 shows L($^{13}$CO)-L$_{IR}$ and L(C$^{18}$O)-L$_{IR}$ relations in the log-log plot. The solid lines are the fit results for all GMCs. The least-squares fit results are also noted in each panel. The correlation between $^{12}$CO and IR luminosities is not given in Figure 4, but rather only the least-squares fit result is given below.

$$\log(L_{IR}) = 0.41 \times \log(L_{^{12}CO}) + 3.02, (c.c. = 0.40)$$

At first glance of Figure 4, the correlations between the IR luminosity and $^{12}$CO, $^{13}$CO and C$^{18}$O luminosities are present but with large scatter. The L$_{IR}$-L($^{13}$CO) and L$_{IR}$-L(C$^{18}$O) correlations are much tighter than the L$_{IR}$-L($^{12}$CO) correlation with steeper slopes. Compared with statistical results of the L($^{12}$CO)-L$_{IR}$ relation for the high luminosity GMCs sample (L$_{CO}$ ~$10^3$—$10^6$ K km s$^{-1}$pc$^2$, L$_{IR}$ ~ $10^4$—$10^7$L$_\odot$ ) of Mooney & Solomon (1988), their results are closer to being a linear correlation with a slope of 0.96, but the scatter is also large. Note that a very flat correlation is also presented in the sample of Mooney & Solomon if L$_{IR}$<$10^{5.8}$ L$_\odot$ (their Fig. 1). This might imply that on the GMC scale, the IR luminosity as an SFR indicator of high luminosity and low luminosity GMCs is different. We know that in small molecular clouds, IR luminosity may not be a good SFR indicator since the interstellar radiation fields (ISRF) can significantly contribute to L$_{IR}$.

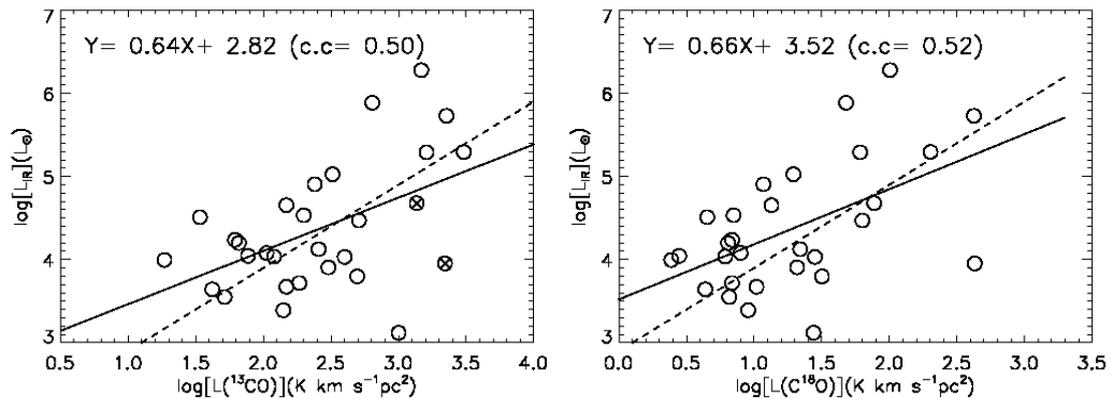

Figure 4. Correlations of $^{13}$CO-IR (left) luminosities and C$^{18}$O-IR (right) luminosities in the log-log plots. The solid lines are the least-squares fit results; the fit results and correlation coefficients are noted in each plot. The dashed lines with slope of 1 are Y=1.00X+1.90 and Y=1.00X+2.90, respectively.

3. Discussion of CO isotopic luminosity ratios, CO isotopic and IR luminosity ratios, and SFR and SFE.

Gao & Solomon (2004) found a linear correlation between HCN and IR luminosities for 65 external star-forming galaxies, which is $\log L_{IR} = 1.00 \log L_{HCN} + 2.9$ [6,7], with correlation coefficient R=0.94. Subsequently, Wu et al. (2005) also revealed a similar linear correlation between HCN and IR luminosities for 27 of 47 GMCs with IR luminosities larger than $10^{4.5} L_\odot$, which is $\log L_{IR} = 1.02 \log L_{HCN} + 2.79$ [10], with a larger scatter than the $L_{IR}$-$L_{HCN}$ relation on a galactic scale. Many of the low luminosity GMCs in the sample of Wu et al. obviously depart from a linear correlation and show a larger dispersion. On the GMC scale, we discussed in last section that the CO-IR correlation for our sample mapped in CO J=1-0 isotopes is significantly different from that of the higher luminosity GMCs sample of Mooney & Solomon (1988), which is basically linear, not to mention the HCN-IR relation for GMCs or galaxies. Only 9 GMCs in our sample belong to high luminosity sources (with $L_{IR} > 10^{4.5} L_\odot$), thus it is not feasible here to give a fit result for the high luminosity sources alone. As a whole, the HCN-IR relation established in galaxies even in dense cores of GMCs shows tighter correlation than the $^{12}$CO-IR, $^{13}$CO-IR and $C^{18}$O-IR relations. One obvious reason for this trend is that our sample is dominated by low luminosity GMCs (~70%). The low luminosity GMCs in the sample of Wu et al. also obviously depart from a linear correlation and show larger scatter. Wu et al. (2005) suggested that GMCs with $L_{IR} < 10^{4.5} L_\odot$ belong to small cores, and they might share a different initial mass function with GMCs of higher IR luminosity. Another possible explanation is that the ISRF is more significant in small clouds, which has been mentioned in the last section. A large and complete sample is useful for confirming this conclusion.

The two panels of Figure 5 present the distance independent $L(^{13}CO)$-$L_{IR}$ and $L(C^{18}O)$-$L_{IR}$ correlations normalized by $L(^{12}CO)$ in the log-log plot, respectively. The $L(^{13}CO)/L(^{12}CO)$ and $L(C^{18}O)/L(^{12}CO)$ ratios are indicators of gas that are traced by $^{13}CO$ and $C^{18}O$ to the total gas traced by $^{12}CO$. From $^{12}CO$ to $^{13}CO$ to $C^{18}O$, the traced column density of gas increases with decreasing optical depth of molecular lines, and $C^{18}O$ traces the highest column density. In Figure 5, on the x-axis, $L(^{13}CO)/L(^{12}CO)$ and $L(C^{18}O)/L(^{12}CO)$ indicate the molecular gas fraction at high column density, and on the y-axis, $L(^{12}CO)/L_{IR}$ indicates the SFE. The weak correlations (dashed lines) suggest that SFE is proportional to the molecular gas fraction at high column densities. The solid lines in figure 5 are the least-squares fit results for all of the 29 GMCs, whereas the dash lines are the fit results for only 27 of 29 GMCs after excluding the two GMCs (G35.20-0.74 and G24.49-0.04) with low signal-to-noise ratios. Comparing the solid line with the dashed line, it is obvious that the fit result is strongly affected by these two outlier sources with large deviations (these two points in most case are the most extreme outliers in previous plots, e.g. Figs. 2 and 4). The effects are more obvious in the $^{13}$CO-IR relation, which almost results in the disappearance of the correlation (C.C.=0.15) if we include the two GMCs with low signal-to-noise ratios.

Figure 5 shows that the $^{13}$CO-IR and $C^{18}$O-IR correlations might be slightly different. The left panel shows larger scatter than the right panel, though both correlations are very weak. In other words, $C^{18}$O-IR shows a tighter correlation than $^{13}$CO-IR normalized by $L(^{12}CO)$, even though the

$^{13}$CO-IR and C$^{18}$O-IR correlations are essentially same (Figure 4). Evidently, these correlations in dense cores of GMCs show larger scatter and much weaker correlations than the HCN-IR relation normalized by L($^{12}$CO) in the external galaxies sample of Gao & Solomon (2004). Although $^{13}$CO and C$^{18}$O are optically much thinner than HCN, the critical density of HCN (~$10^5$cm$^{-3}$) is much higher than that of $^{13}$CO and C$^{18}$O which is only about $10^3$cm$^{-3}$. $^{13}$CO and C$^{18}$O can trace the high column density region of the molecular cloud, but not necessary at an extremely high volume density, which seems to not be closely related to star formation. This might be interpreted in that both the SFR and SFE are mainly determined by the molecular gas at high local volume density rather than high column density. Moreover, this also suggests that the column density is not well correlated with the volume density for the 29 GMCs in our sample.

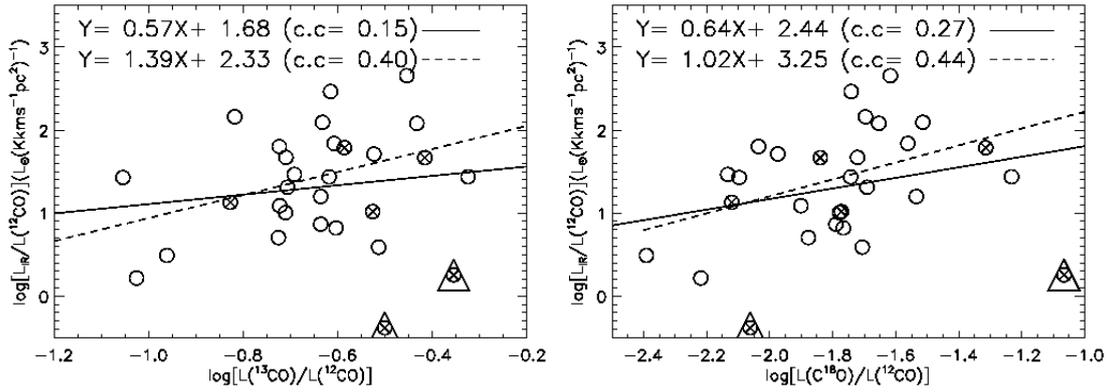

Figure 5. Correlations of L($^{13}$CO)-L$_{IR}$ (left) and L(C$^{18}$O) and L$_{IR}$ (right) normalized by L($^{12}$CO) in the log-log plots. The solid and dashed lines indicate the fit results before or after excluding the sources marked with open triangles. The fit results and correlation coefficients are noted in each panel.

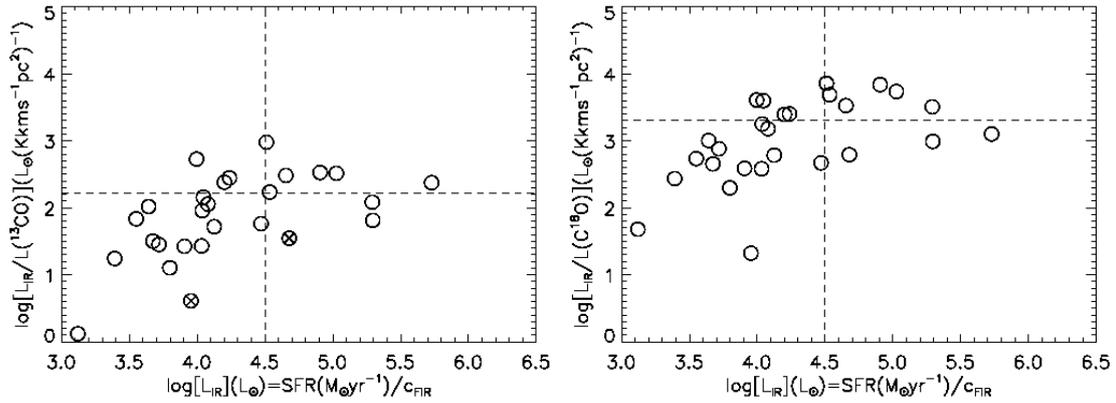

Figure 6. Left panel: Correlation between L$_{IR}$ and L$_{IR}$/L($^{13}$CO) in the log-log plot; Right panel: Correlation between L$_{IR}$ and L$_{IR}$/L(C$^{18}$O) in the log-log plot. The dashed lines mark the mean value of L$_{IR}$/L($^{13}$CO), L$_{IR}$/L(C$^{18}$O) and the cutoff luminosity L$_{IR}$=$10^{4.5}$L$_\odot$ which was defined by Wu et al. (2005)[10], respectively.

The IR and CO luminosity ratios are regarded as SFE indicators. The left and right panels of Figure 6 show L$_{IR}$ of all 29 GMCs in the sample as a function of the L$_{IR}$/L($^{13}$CO) and L$_{IR}$/L(C$^{18}$O) ratios in the log-log plot, respectively. $L_{IR}$ and $L_{IR}/L(^{12}CO)$ have similar relations in the log-log plot, and many studies have shown such a similar plot. For the 29 GMCs, $<L_{IR}/L(^{12}CO)> = 58$ $L_{\odot}(Kkms^{-1}pc^2)^{-1}$, with median value of 27 $L_{\odot}(Kkms^{-1}pc^2)^{-1}$; $<L_{IR}/L(^{13}CO)> = 237$ $L_{\odot}(Kkms^{-1}pc^2)^{-1}$, with median value of 104 $L_{\odot}(Kkms^{-1}pc^2)^{-1}$; $<L_{IR}/L(C^{18}O)> = 3107$ $L_{\odot}(Kkms^{-1}pc^2)^{-1}$, with median value of 1266 $L_{\odot}(Kkms^{-1}pc^2)^{-1}$. The mean values of $L_{IR}/L(^{13}CO)$ and $L_{IR}/L(C^{18}O)$ are both marked in Figure 6. Note that the $<L_{IR}/L(^{12}CO)>$ of the active GMCs sub-sample of Mooney & Solomon (1988) is ~3 times smaller, since our CO isotope observations are focused on the emission from dense cores of GMCs rather than the whole GMCs. Wu et al. (2005) pointed out that the mean value of $L_{IR}/L_{HCN}$ is around 900 $L_{\odot}(Kkms^{-1}pc^2)^{-1}$ and is independent on $L_{IR}$, from galaxies to GMCs, as long as $L_{IR}>10^{4.5}$ $L_{\odot}$. But a decline in $L_{IR}/L_{HCN}$ ratio occurs when $L_{IR}<10^{4.5}$ $L_{\odot}$[10]. This has been interpreted as indicating that HCN emissions can trace the fundamental unit of star formation: different galaxies have different counts using such a unit. In cores with $L_{IR}<10^{4.5}$ $L_{\odot}$, the fraction of newly formed massive stars rises rapidly with the increase of core mass, so the SFE is not a constant [10]. Recent theoretical work shows that besides HCN, any molecular line with $n_{cri} \geq <n>$ is expected to have such kind of property[15, 16]. From our limited sample of GMCs, we can also conclude that there are similar trends in the high luminosity case for which $L_{IR}>10^{4.5}$ $L_{\odot}$. The $L_{IR}/L(^{12}CO)$, $L_{IR}/L(^{13}CO)$ and $L_{IR}/L(C^{18}O)$ ratios are almost independent of $L_{IR}$; but $L_{IR}/L(^{12}CO)$, $L_{IR}/L(^{13}CO)$ and $L_{IR}/L(C^{18}O)$ are proportional to $L_{IR}$ if $L_{IR}<10^{4.5}$ $L_{\odot}$. We note that among $^{12}CO(1-0)$, $^{13}CO(1-0)$ and $C^{18}O(1-0)$, it is almost impossible for $^{12}CO$ to have the property of $n_{cri} \geq <n>$ due to the nature of its low critical density. Thus, similar trends between $^{12}CO$ and HCN might be interpreted as selection effects, etc. $^{13}CO$ and $C^{18}O$ have relatively higher critical densities than $^{12}CO$. Therefore, $^{13}CO$ and $C^{18}O$ might be similar to HCN, when $L_{IR}<10^{4.5}$ $L_{\odot}$, that is, the fraction of newly formed massive star is proportional to the core mass. In other words, the SFE is proportional to the core mass. Nine dense cores in our sample with $L_{IR}>10^{4.5}$ $L_{\odot}$ show constant SFE. We also caution that $L_{IR}$ at such low vales ($L_{IR}<10^{4.5}$ $L_{\odot}$) could be significantly contributed by the ISRF and it might not be a reliable tracer of SFR. A statistically significant sample similar to Mooney & Solomon (1988) is needed to completely explore the full range of properties of the dense cores and, thus, is useful to confirm or refute the trends found in this study.

4. Summary of the results.

Combining the $^{12}CO$, $^{13}CO$, $C^{18}O$ data with the IRAS four band data, here, we estimate the physical parameters for 29 dense cores in GMCs. This paper analyzes the correlations between CO isotopic luminosities and infrared luminosity. Furthermore, the relationships between the molecular gas tracers and star formation rate are discussed and compared with the HCN-IR correlations of 47 dense cores in the Galaxy and 65 external star-forming galaxies. The main results in this paper are summarized below:

(1) We estimate the physical parameters such as size, viral mass, and CO J=1-0 isotopic and infrared luminosities, which were given in the form of mean and median values. All of these parameters are comparable to the typical values in high mass star-forming regions. The derived values of $^{12}CO$ show the highest and the values of $C^{18}O$ show the lowest.

(2) Tight correlations among $^{12}CO$, $^{13}CO$ and $C^{18}O$ luminosities are present. $L(^{12}CO)$-$L(^{13}CO)$

shows the tightest relation and L($^{12}$CO)-L($C^{18}$O) shows the weakest relation, which might be significantly determined by the size of the GMC, the optical depth and the excitation of the molecular line.

（3）The correlations between IR luminosity and CO isotopic luminosities are present. The least-squares fit results for $^{12}$CO, $^{13}$CO and $C^{18}$O dense cores are $\log(L_{IR}) = 0.41 \times \log(L_{^{12}CO}) + 3.02, (c.c. = 0.40)$, $\log(L_{IR}) = 0.64 \times \log(L_{^{13}CO}) + 2.82, (c.c. = 0.50)$ and $\log(L_{IR}) = 0.66 \times \log(L_{C^{18}O}) + 3.52, (c.c. = 0.52)$, respectively. The correlations of L($^{13}$CO)-L$_{IR}$ and L($C^{18}$O)-L$_{IR}$ are similar and better than L($^{12}$CO)-L$_{IR}$. After being normalized by $^{12}$CO luminosity, the weak correlation of L ($C^{18}$O)-L$_{IR}$ $(c.c. = 0.27)$ is a bit tighter than the weak correlation of L($^{13}$CO)-L$_{IR}$ $(c.c. = 0.15)$. Compared with the L$_{HCN}$-L$_{IR}$ relation normalized by CO[7, 8], these correlations show larger scatter and are much weaker. This might be interpreted as both the SFR and SFE being mainly determined by the molecular gas at high volume density rather than high column density.

In short, the comparison of CO J=1-0 isotopic and IR luminosities of 29 dense cores led to some results summarized above. Unlike $^{13}$CO, the $C^{18}$O mapping in dense cores of GMCs is still very limited[19]. We need more $C^{18}$O observations to improve the statistics. We also expect that the combination of higher resolution far-IR (eg. Spitzer and Herschel) and high-J CO, such as the CO(3-2) warm dense gas tracer [20, 21] observations can provide us more useful information to better reveal the active SF property of these dense cores.

Acknowledge    The authors thank the help from the staff at the Qinghai station, Purple Mountain Observatory. We also appreciate the referees' useful suggestions very much and thank Ms. Zhang, Li-Yun for providing some of the raw CO data.